# An Adaptive Decision-Making Approach for Better Selection of a Blockchain Platform for Health Insurance Frauds Detection with Smart Contracts: Development and Performance Evaluation

Rima Kaafarani[a], Leila Ismail[b,c,d*], Oussama Zahwe[a]

[a] *ICCS-Lab, Computer Science Department, American University of Culture and Education, 1507 Beirut, Lebanon*
[b] *Cloud Computing and Distributed Systems (CLOUDS) Laboratory, School of Computing and Information Systems, The University of Melbourne, Melbourne, Australia*
[c] *Intelligent Distributed Computing and Systems (INDUCE) Laboratory, Department of Computer Science and Software Engineering, College of Information Technology, United Arab Emirates University, Abu Dhabi, United Arab Emirates*
[d] *National Water and Energy Center, United Arab Emirates University, Abu Dhabi, United Arab Emirates*
*\*Correspondence: Leila Ismail (email:leila@uaeu.ac.ae)*

**Abstract**

Blockchain technology has piqued the interest of businesses of all types, while consistently improving and adapting to developers and business owners requirements. Therefore, several blockchain platforms have emerged, making it challenging to select a suitable one for a specific type of business. This paper presents a classification of over one hundred blockchain platforms. We develop smart contracts for detecting healthcare insurance frauds using two blockchain platforms selected based on our proposed decision-making map approach for the selection of the top two suitable platforms for healthcare insurance frauds detection application, followed by an evaluation of their performances. Our classification shows that the largest percentage of blockchain platforms could be used for all types of application domains, and the second biggest percentage is to develop financial services only, even though generic platforms can be used, while a small number is for developing in other specific application domains. Our decision-making map revealed that Hyperledger Fabric is the best blockchain platform for detecting healthcare insurance frauds. The performance evaluation of the top two selected platforms indicates that Fabric surpassed Neo in all metrics.

## 1. Introduction

Frauds in health insurance is the intentional deception or misrepresentation in order to obtain unauthorized benefits to the policyholder, another party, or the entity providing services. The global sustainability of the healthcare system is threatened by the high cost incurred due to healthcare insurance frauds. Consequently, the need for an automated system to detect frauds without human intervention is critical [1]. Machine learning solutions for detecting frauds rely on datasets to train models, which raises privacy and security concerns [2]. Blockchain, a peer-to-peer distributed ledger that is secured with hash functions and public/private keys cryptography, has demonstrated its potential in the healthcare sector by combining important characteristics such as immutability, transparency, and traceability [3]-[9]. Furthermore, blockchain provides the possibility to run smart contracts. Therefore, in this paper, we develop a blockchain-based smart contract for the detection of healthcare insurance frauds. In addition, we present a classification for 102 blockchain platforms and then we devise a methodology to select the best platform that is suitable for the development of healthcare insurance frauds detection. To our knowledge we are the first to provide these solutions for healthcare insurance.

Our contributions in this paper are as follows:

- We develop the detection of frauds by implementing smart contracts, based on a taxonomy of frauds and scenarios.
- We propose a decision-making map to identify the best blockchain platform(s) adequate for the implementation of healthcare insurance frauds detection, and we apply this map to 102 platforms.
- We develop and implement the detection of frauds on the top two selected platforms by the decision-making map, and evaluate their performance.

2) we provide an overview of related work. 3) we present a classification of current blockchain architectures. 4) we discuss our proposed decision-making map for selecting a suitable blockchain platform for healthcare insurance frauds detection. 5) we present the smart contract algorithms for the detection of frauds in healthcare insurance. 6) we present a performance analysis of the top two selected platforms. 7) we conclude the paper.

## 2. Related Work

Healthcare frauds has significantly increased losses for individuals, businesses, and governments. Combating healthcare frauds has become a critical concern. As a result, recent work proposed solutions based on blockchain to detect frauds. Authors in [10] proposed a solution based on blockchain to detect if the claim meets the relevant provisions of the healthcare insurance policy. In [11] proposed a solution to prevent health insurance frauds using two frauds scenarios. Authors in [12] used the Ethereum blockchain to develop a framework to record claims data and transaction patients to be validators to help detect frauds. However, none of these works consider all the frauds scenarios and the performance of blockchains, and the selection of used blockchain platforms is not justified. Therefore, authors in [1] solve this problem by proposing a new blockchain architecture for better performance and presenting a taxonomy of 12 frauds scenarios that implement using smart contracts on two blockchain platforms selected by our decision-making map. Table 1 presents the strengths and weaknesses of recent works using blockchain platforms for healthcare insurance frauds detection.

Table 1. Strengths and weaknesses of works on blockchain-based healthcare insurance frauds detection

| Work | Throughput | Latency | CPU usage | Memory | Number of frauds scenarios implemented | Platform | Reason for choosing platform |
|---|---|---|---|---|---|---|---|
| [10] | X | X | X | X | 1 | Ethereum | N/A |
| [11] | X | X | X | X | 2 | BigchainDB | N/A |
| [12] | X | X | X | X | 3 | N/A | N/A |
| Our work | ✓ | ✓ | ✓ | ✓ | 12 | Hyperledger Fabric & Neo | Based on our proposed decision-making map for healthcare insurance frauds detection |

## 3. Classification of current blockchain architectures

Blockchain is a secure distributed ledger across a network of peers to record transactions without the need for central coordination. Peers are given the choice to approve changes and commit to the ledger through a consensus mechanism [3]. Blockchain has four types, permissionless, permissioned, hybrid, and consortium [3]. Considering the important characteristics of blockchain, many blockchain platforms have emerged for different application domains [8]-[9]. Blockchain platforms enable the development of applications. We classify One Hundred Two platforms based on the application domains they support, Table 2 consists of platforms used for all types of domains, Table 3 constitutes of those which are used for financial services and other specified domains.

Table 2. Blockchain platforms used for all types of domains

| Application domain | Platforms |
|---|---|
| All domains | Fantom Opera Platform, Ignite, Blockchain Service Network (BSN), Quorum, Kaleido, Chain Core, CCF (previously coco), ICON, Elements Block Platform, Chain33, Multichain, Hyperledger Fabric, Tendermint, EOS, Hyperledger Iroha, Stratis Platform, Universa, Secrete Network, Monax, Hyperledger Burrow, Autonity, Eris:db, Qasis, XinFin-XDC network, Ethereum, Hyperledger Sawtooth, Neo, Azure blockchain workbench, BigchainDB, Oracle blockchain, Credits blockchain, Nem, Smilo, LISK, Neblio, Dragonchain, Hyperledger Besu, ÆTERNITY, Cosmos, Near, Cypherium, QANplatform, IBM Blockchain Platform, Hedera hashgraph, Polygon, Algorand, IOTA, Waves, ZILIQA, Elrond, WanChain, Rubix, Stack, CORTEX, Nervos CKB, Casper Network, Open Zeppelin, iOlite. |

Table 3. Blockchain platforms for financial services and other specified domains

| Application domain | Platforms |
|---|---|
| Financial services | TRON, ripple, Monero, Binance Smart Chain, Bitcoin, Litecoin, NXT, Nano, SpaceMint. Parity, Exonum, Bitshares, Stellar, Cordano, Libra, Counterparty, Shadow coin, Decentralized Credits (Decred), Chia, Avalanche, Komodo, Symbiont Assembly, Kadena, FISCO BCOS, R3 Corda |
| Digital identities | Hyperledger Indy |
| Storing structured data | Factom |
| Add Ethereum smart contracts to Bitcoin's blockchain | Qtum |
| Digital asset management | brustCoin, Openchain |
| Blockchain interoperability | Polkadot, Aion |
| DNS | Namecoin |
| Economic | Vechain |
| Global trade | CargoX |
| IoT | Tikiri |
| Metaverse | Klaytn |
| Money transfers | Peercoin |
| NFT, DeFi, payment, gaming, DAO | Solana |
| Ownership | Poex.io |
| Security for IoT | Iotchain |
| Social media | Steem.it |
| Storage | Byteball, Filecoin |

## 4. Decision-Making Map for Healthcare Frauds Detection Platform

Due to the proliferation of blockchain platforms, it becomes a very tedious task to select a platform for development. [13] proposed a methodology for selecting a platform, however, it considers category features that may not be needed by a particular application and does not consider the specificity of healthcare insurance frauds detection. Therefore, we propose a decision-making map to select a platform for healthcare insurance frauds detection. It considers three main categories of blockchain features, "compulsory features", "recommended features" which are significantly adequate features, and "possible features" that would be preferably existent in the platform. Categories are given importance weights. The first category consists of seven features: 1) The application layer enables to build user interface and run smart contracts for health insurance. 2) Network layer allows a peer-to-peer decentralized network. 3) Protocol layer specifies the consensus protocol; we use Byzantine-based consensus because it is lightweight in terms of memory and CPU usage [3]. 4) Interoperability technologies such as Oracle, to import data for smart contracts from off-chain resources. 5) On-chain transaction where a transaction is executed on the main blockchain for security, decentralization, and transparency. 6) Permissioned blockchain, where the ledger is shared only with trusted nodes. 7) Smart contracts to execute algorithms for detecting healthcare insurance frauds. The second category consists of six features: 1) Enterprise system interrogation provides easy access to data, seamless data flow, and time and cost savings. 2) Private so that only authenticated users can participate in the network. 3) Turing complete indicates that the blockchain platform's virtual machine can solve any computational problem. 4-5-6) we select JavaScript, Python, and Solidity because they are easy to use and learn. Finally, in the third category, we consider seven features: 1-2) Java and Golang, they share the same reason as the last 3 features in the second category of the "recommended features" list. 3) Virtual Machine to run smart contracts. 4) Data privacy technology and certificates for authentication are essential when dealing with sensitive patient data. 5) Zero-knowledge proof which is an encryption scheme in which one party (prover) guarantees another party (verifier) that they know value X without revealing the actual value. 6) Cryptographic tokens, could be used for payments. 7) Cross-chain interoperable to link two independent blockchains. Our results show Hyperledger Fabric [14] as the most suitable platform, followed by Neo [15], then Quorum, Hydrachain, and Chain. R3 Corda, and BigchainDB are eliminated because they missed some of the features in the "compulsory features" category. Fig. 1 shows the percentage score for the top five platforms that are deemed suitable.

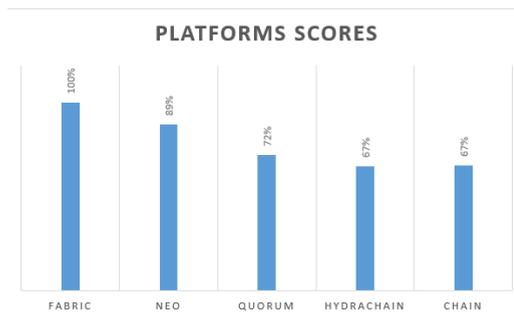

Fig. 1 Platforms scores

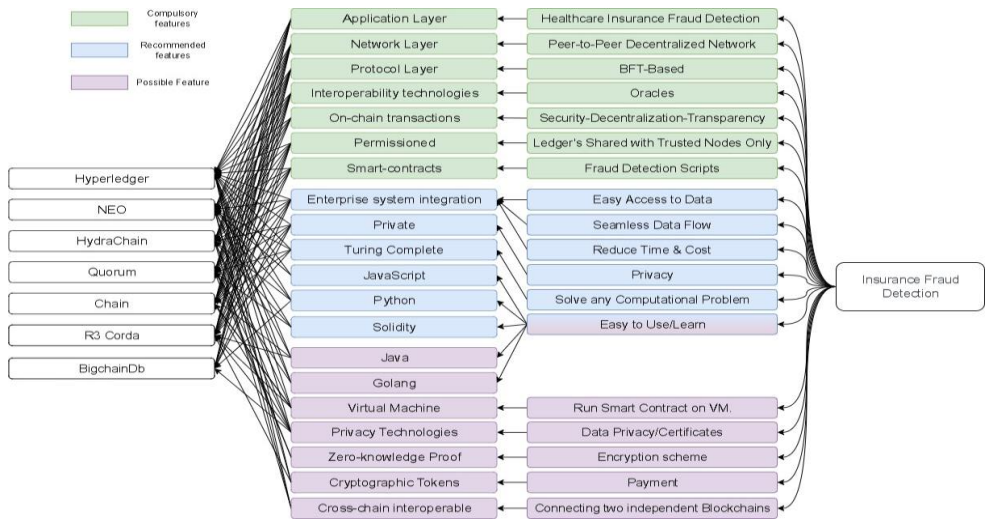

Fig. 2. Proposed decision-making map for selecting a platform for healthcare insurance frauds detection

Fig. 2 shows our adaptive-blockchain-frauds detection-based selection map, with examples of how we map the platforms to select the best platforms for fraud detection development. We apply this mapping to all the platforms used for all types of application domains to find their suitability for fraud detection, after eliminating the platforms that use permission-less blockchains and do not support smart contracts. As a result, we apply the map on 46 platforms.

## 5. Algorithms for Smart-Contract frauds detection

We focus on verifying submitted claims to determine whether or not they are fraudulent. For each fraud scenario, we look for a detectable pattern. In some scenarios we need data from external sources, for example in Table 4, algorithm 1 requires a list of the licensed healthcare providers, which can be obtained from the ministry of health. Algorithm 2 covers three fraud scenarios in one function, the pattern that should be found is the same: 1) when a healthcare provider is referring patients within the same healthcare provider, 2) involves a provider referring patients to earn a commission, and 3) involves the providers having a financial relationship with the healthcare provider to whom they refer patients (see Appendix). The participants in the network are allied health professionals, patients, pharmacists, diagnostic centers, pharmaceutical companies, the branch medical insurance company, the head medical insurance company, hospitals, and medical equipment suppliers.

Table 4. Algorithms for smart-contract frauds detection

| Algorithm 1: UnlicensedCareProvider | Algorithm 2: CheckForReferring |
|---|---|
| 1  begin | 1  begin |
| 2  get claim data from ledger using claim id | 2  get claim data from ledger using claim id, healthcare provider id |
| 3      if claim does not exist | 3  if claim id does not exist |
| 4          return " claim does not exist" | 4      return "claim does not exist" |
| 5  put claim data in struct using json (healthcare provider id, invoice id) | 5  put claim data (provider id, invoice id) in struct using json |
| 6  get invoice from ledger using invoice id from the claim data | 6  create a list for providers invoices |
| 7  if the invoice exists | 7  get all invoices from ledger (provider id, invoice id, patient id) |
| 8      return "Invoice does not exist" | 8  for each invoice in the invoices, we've got from the ledger |
| 9      end | 9      if provider from the claim == provider in the invoice |
| 1  put invoice data in struct using json (provider id) | 10         put the invoice id in the list of providers invoices |
| 11 get the healthcare provider from the invoice data | 11 end for |
| 12 get a list of the licensed providers | 12 loop over the providers' invoice IDs |
| 13 for each provider in licensed providers list | 13     put providers names in a list |
| 14     if provider exist | 14 end |
| 15         return "not fraud" | 15 set counter =0 |
| 16     end | 16 for each provider id in providers list |
| 17 return "fraud" | 17     if provider id is found |
| 18 end | 18         counter ++ |
| | 19 if counter = threshold |
| | 20     check if the provider in claim is the same provider in the invoices |
| | 21         return "fraud, pining the system" |
| | 22     end |
| | 23     else check for financial relation between provider in the claim and in the invoices |
| | 24         return "fraud, financial relation is found" |
| | 25     end |
| | 26     else check if provider taking commission |
| | 27         return "fraud, provider taking commission" |
| | 28     end |
| | 29     else return "threshold is reach but no financial found or taking commission, no fraud" |
| | 30 end |
| | 31 end |
| | 32 return "Not a fraud" |
| | 33 end |

## 6. Performance Evaluation

### 6.1. Experiments

We develop smart contracts algorithms for the fraud scenarios that were identified in [1], using the two platforms which have been selected as the top two by our decision-making map. A transaction includes one record of the claim and requires files for verification e.g., invoices. We evaluate the performance of the platforms using throughput, latency, CPU usage, and Memory usage performance metrics. We set two testing scenarios, the first one consists of sending a constant number of transactions by peers for a duration ranging from 30sec to 120sec. In the second scenario, we increase the number of transactions sent over the network from 1000TX to 10,000TX.

### 6.2. Experimental Environment

We used Ubuntu 20.04 WSL2 on Windows 11 operating systems. Fabric was running on Docker, with all peers joining one channel representing a single insurance company. Nodes run on Fabric v2.2, with 4 orgs and one orderer. we use a batch size of 500. Neo private blockchain was created using the N3 Neo Visual DevTracker extension on Visual Studio Code. We used c# as the programing language recommended in the documentation of Neo, and use a batch size of 500 as well. On both platforms, we use LevelDB which is a key value data storage. For the hardware setup, we utilize a 16 GB Ram, Core i7 11th gen 2.80 GHz processor. For the benchmarking, we use Hyperledger Caliper [16] for Fabric, and Neo-bench [17] for Neo.

*6.3. Results analysis*

Fig. 3 demonstrates the throughput for Neo and Fabric in different scenarios. In the first scenario we set a fixed transaction number to be sent for a duration in seconds Fig. 3 (a, b), Neo increases at the duration of 60 seconds and reaches a throughput of 629 transactions per second (TPS), but then declined to 304 TPS and held to that to the end of the test. On the other hand, Fabric showed a similar pattern, it started with a high throughput of almost 800 TPS then decreased to ~450TPS at the end of the test. In The second test scenario with an increasing number of transactions, Fabric outperformed Neo again in this test and was increasing with the number of transactions sent, while Neo was showing fluctuations. As for latency, Neo always takes from 14 to 24 sec to confirm a transaction.

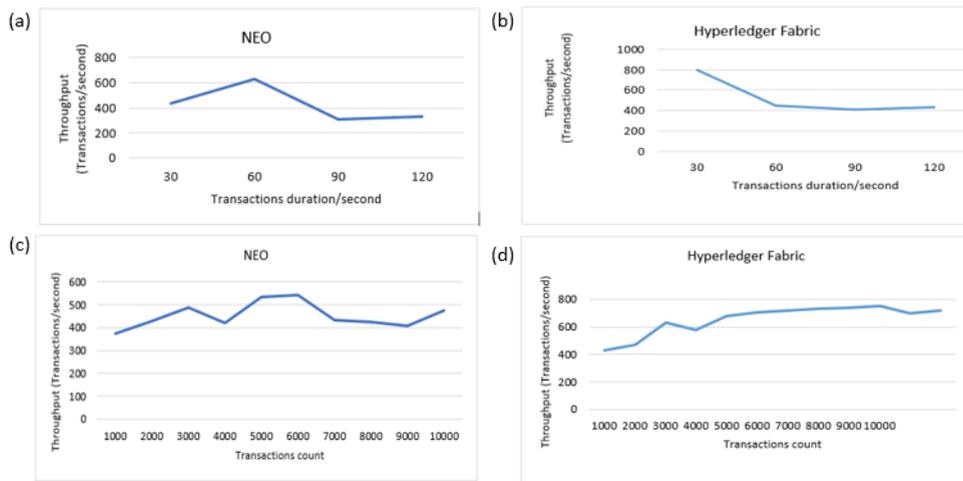

Fig. 3. Throughput for (a) Neo while sending 1000TX for a certain duration. (b) Hyperledger Fabric throughput while sending 1000TX for a certain duration. (c) Neo with transaction count ranging from 1000TX to 10,000TX. (d) Hyperledger Fabric with transaction count ranging from 1000TX to 10,000TX.

Fig. 4 (a) shows the Fabric's latency during the second test scenario; although latency increased slightly, it remained between 0.03sec and 0.04sec. Fig. 4 (b) displays the Fabric's latency for the first test scenario, latency increases with more transactions being submitted.

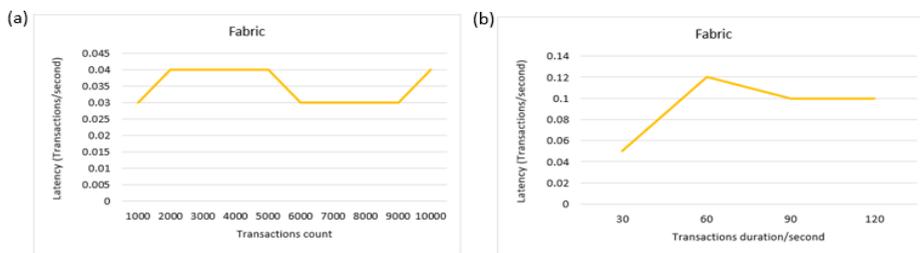

Fig. 4. Fabric latency result (a) sending transactions from 1000TX to 10,000TX. (b) sending 1000TX for a certain duration.

Fig. 6 shows the CPU and memory usage for the platforms, Fabric used fewer resources than NEO. During all the tests, the two platforms score a 100% success rate, making them more reliable platforms for the sharing of sensitive healthcare insurance data. Fabric outperformed Neo in all the tests, demonstrating that it is the best choice for creating healthcare insurance frauds detection.

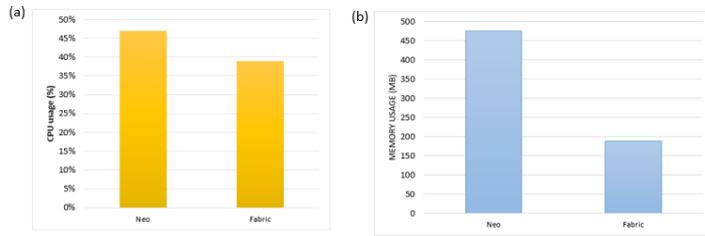

Fig. 5. (a) Memory usage of Fabric and Neo. (b) CPU usage for Fabric and Neo

## 7. Conclusion and Future Work

The number of blockchain platforms is growing at an exponential rate, and software companies are becoming increasingly interested in incorporating blockchain technology into their products. Medical insurance frauds detection is a significant issue in the medical industry. Every year, healthcare insurance frauds costs tens of billions of dollars, and numerous frauds endanger patients' health. In this paper, we develop smart contracts for the detection of healthcare insurance frauds, which is based on the taxonomy of frauds scenarios in [1]. In addition, we propose a decision-making map, which allows us to identify the best blockchain platform for healthcare insurance frauds detection. Finally, we develop and implement the detection of frauds on the top two selected platforms by the map, and evaluate their performance; Fabric outperformed Neo; Fabric's network structure and features are far better suited to the use case, and its performance could be enhanced with using configurations options.

## Appendix

**Algorithm 3: DoctorShopping**
```
1   begin
2     get claim data (patient id)
3     get last five invoices for the patient (date, provider id, drugs list)
4     get dates from invoices
5     get IDs of providers
6     check if dates are sorted to check if the patient is buying the medications on a regular basis
7       get the difference between dates to make sure there is two months maximum between purchases
8     end
9     if date difference between dates are less than two months
10      get providers IDs
11      check if provider id is repeated
12        if provider id is not repeated more than two times
13          return "patient buying the medication regularly but visiting the same provider more often, not fraud"
14        end
15      else
16        return "Buying meds regularly and from multiple providers is detected, fraud"
17      end
18  end
```

**Algorithm 4: Equipment**
```
1   // Increasing price of medical equipment for patients whose insurance
    covers equipment costs or claiming for expensive equipment while
    providing the ones with les expense
2   begin
3     get claim data (claim id, invoice id, amount paid, equipment)
4     get the amount paid by the patient from the claim
5     get the equipment details
6     get price list for the equipment from other resources
7     for each price in the price list
8       if price > max
9         max = price
10      end if
11    end for
12    if amount paid > max
13      return "fraud"
14    end if
15    else
16      return "fraud"
17    end if
18  end
```

**Algorithm 5: CommisionBased**
```
1   // Providing specific brand of medicines to get a
    commission from the pharmaceutical company
2   begin
3     get claim data (provider id)
4     get fraudster id from the claim
5     get all prescription (provider id, med list)
6     for each prescription
7       if fraudster id exists
8         put prescribed medications in a list
9       end if
10    end for
11    for each medication in list
12      count repeated meds
13      if counter reached threshold
14        check if the fraudster is taking commission
15        if true
16          return "fraud"
17        end if
18      end if
19    end for
20    return "not fraud"
21  End
```

**Algorithm 6: ManagedCare**
```
1   // Denial of service, substandard care and
    creation of administrative obstacles for patients
    by the managed care organizations
2   begin
3     get claim data
4     get all invoiced where date = date in claim
5     get fraudster invoices
6     if other patient has records at that date
7       counter ++
8     end if
9     if counter== threshold
10      return "fraud"
11    end if
12    else
13      return " not fraud"
14    end else
15  end
```

**Algorithm 7: ManipulateDiagnosis**

```
1   // Manipulation of diagnosis in the claims without
    the knowledge of the patients
2   begin
3   get claim data (claim id, diagnosis code)
4   get invoice id from claim
5   get invoice (diagnosis code)
6   invDiagnosis= diagnosis code from invoice
7   claimDIagnosis = diagnosis code from claim
8   if onvDiagnosis!=claimDiagnosis
9       return "fraud"
10      end if
11  else
12      return " not fraud"
13      end else
14  end
```

**Algorithm 8: UnapprovedOffLable**

```
1   // A pharmaceutical company provides
    incentives to doctors to promote unapproved or
    off-label drugs
2   begin
3   get claim data (claim id, provider id)
4   get prescription for the fraudster
5   get drugs from prescriptions
6   get list of approved and labeled drugs
7   for each drug in the list
8       if drug ! exist in list
9       return "fraud"
10      end for
11  return " not fraud"
12  End
```

**Algorithm 9: UnwantedCare**

```
1   // Providing unwanted care to the patients,
    increasing service hours in the bill, duplicate
    claims, phantom billing, or replacing code of
    diseases with ones with more price
2   begin
3   get claim data (claim id, invoice id, provider id)
4   check if claim exists with same data
    CheckClaim(id)
4   get invoices for the fraudster
5   get diagnosis
6   ask for another doctor opinion
7   if doctor replay positive
8       return "fraud"
9       end if
11  return " not fraud"
12  end
```